\documentclass[letterpaper]{IEEEtran}

\usepackage[utf8]{inputenc}
\usepackage[cmex10]{amsmath}
\usepackage{amsfonts}
\usepackage{amssymb}
\usepackage{graphicx}
\usepackage{float}
\usepackage{cite}

\graphicspath{ {./figures/} }

\usepackage{url}

\begin{document}
\title{Lighthouse: A User-Centered Web Service\\
        for Linear Algebra Software}

\author{
\IEEEauthorblockN{
Boyana Norris\IEEEauthorrefmark{1},
Sa-Lin Bernstein\IEEEauthorrefmark{2},
Ramya Nair\IEEEauthorrefmark{3}, and
Elizabeth Jessup\IEEEauthorrefmark{3}}

\IEEEauthorblockA{\IEEEauthorrefmark{1}Department of Computer and Information Science, University of Oregon, Eugene, OR, USA}
\IEEEauthorblockA{\IEEEauthorrefmark{2}Computation Institute, University of Chicago and Argonne National Laboratory, Chicago, IL, USA}
\IEEEauthorblockA{\IEEEauthorrefmark{3}Department of Computer Science, University of Colorado, Boulder, CO, USA}
}

\maketitle

\begin{abstract}

Various fields of science and engineering rely on linear algebra 
for large scale data analysis, modeling and simulation, machine learning, and
other applied problems. Linear algebra computations often dominate the
execution time of such applications. Meanwhile, experts in these domains typically
lack the training or time required to develop efficient, high-performance
implementations of linear algebra algorithms. In the Lighthouse project, we
enable developers with varied backgrounds to readily discover and
effectively apply the best available numerical software for their problems. We
have developed a search-based expert system that combines expert knowledge, machine
learning-based classification of existing numerical software collections, and
automated code generation and optimization. 
Lighthouse provides a novel software engineering environment 
aimed at
maximizing both developer productivity and application performance
for dense and sparse linear algebra computations.
\end{abstract}




\section{Introduction}
\label{sec:intro}

Scientists and engineers in a wide variety of disciplines rely on linear algebra algorithms 
(e.g., ~\cite{vidal,Katz,bryan200625}) and their high-performance implementations in 
an ever-expanding field of numerical libraries~\cite{JackMark:Online}.
Selecting a suitable library and using it effectively to solve a given problem generally require 
significant background in numerical analysis, high-performance computing (HPC), and software engineering. It also 
typically involves reading documentation (when available) or researching publications outside of the 
developer's area of expertise. Hence, while continuous advances in numerical analysis and HPC libraries 
allow scientists and engineers to solve larger and more complex problems than ever before, the 
likelihood that the most relevant and best-performing solution method is identified by each 
potential user is steadily decreasing. 

Linear algebra often constitutes the most time-consuming part of scientific applications' execution time; therefore, reducing the
costs of these computations can have a significant impact on overall software performance. 
The process of converting linear algebra from base algorithm to
high-quality implementation, however, is a complex one. Generally, such transformations require expertise in numerical computation,
mathematical software, compilers, and computer architecture and require large, sustained efforts. 
Current high-performance implementations of numerical linear algebra software are based on decades of applied mathematics research
and many thousands of person-hours of effort that is still ongoing. 
Thus, to optimize both scientific productivity and application performance, an application developer whose problems are too large or
complex to solve efficiently through simple sequential algorithms simply cannot afford to develop his or her own 
HPC implementations and \emph{must} use HPC libraries developed by others. 

Matching numerical capabilities with users' specific HPC needs is currently not supported by software engineering tools.
With the Lighthouse project, our goal is to provide a structured taxonomy-based interface to high-performance linear algebra software
and thereby to enable
different types of users to use HPC libraries effectively. We combine taxonomy-based search for identifying specific
solution methods with code generation and optimization capabilities to accommodate a variety of different use cases
that may arise in HPC software development. 

A number of taxonomies exist to aid developers in the translation of linear algebra algorithms to numerical software
(e.g.,\cite{JackMark:Online,blas:Online,GAMS,LAPACKSearch}). However, none of these taxonomies provides an accessible
interface, nor do they supply tools for high quantity code production. Lighthouse~\cite{lighthouse:Online} is the first framework that offers a
searchable ontology of linear algebra software with code generation and tuning capabilities. Like lighthouses 
that aid in maritime navigation,
our Lighthouse will guide practitioners through the obscure maze of numerical software development.
The contributions described in this paper include the following.

\begin{itemize}
    \item Construction of a software taxonomy that provides an organized anthology of software components for linear algebra.
    \item Functionality- and performance-based search of high-performance numerical software capabilities with current support for sequential and parallel dense and sparse linear algebra computations provided by the 
LAPACK~\cite{LAPACK,lapack:Online},
PETSc~\cite{petsc-user-ref,petsc-web-page,petsc-efficient}, and SLEPc~\cite{Hernandez:2005:SSF,slepc:Online}
libraries.
    \item Generation of code templates based on search results to automate initial application design.
    \item Generation of highly tuned code from high-level computation descriptions expressed in a MATLAB-like language. 
\end{itemize}

This paper is organized as follows. Section~\ref{sec:background} presents our motivation and related work.
Section~\ref{sec:method} describes the procedure of developing the Lighthouse taxonomy that covers 
LAPACK, PETSc, and SLEPc, the strategies for designing the user-centered interfaces that offer effective
search functions, and the methods for establishing the support for high-performance implementation.
Section~\ref{sec:result} discusses the specific implementation approaches for the different numerical packages we have integrated into
Lighthouse. Section~\ref{sec:conclusion} presents our conclusions and outlines future work.

\section{Background}
\label{sec:background}

A number of of 
taxonomies exist to aid the code developer in translation of linear algebra 
algorithms to numerical software.  Perhaps the oldest one is the Netlib 
Mathematical Software Repository~\cite{blas:Online}, started in 1985, which 
contains freely available software, documents, and databases pertaining to 
numerical computing including linear algebra.  Contents are provided as 
lists of packages or routines, with or without some explanatory words.  
In newer work, the Linear Algebra Software 
Survey~\cite{SoftwareSurvey} lists over sixty items categorized as support 
routines, dense direct solvers, sparse direct solvers, preconditioners, 
sparse iterative solvers, and sparse eigenvalue solvers together with a 
checklist specifying problem types for each entry.   NIST's Guide to 
Available Mathematical Software (GAMS)~\cite{GAMS} includes even more basic 
linear 
algebra software along with software for a variety of other numerical 
applications.   While the Linear Algebra Software Survey is a linear list, 
GAMS allows search by problem solved, package name, module name, or text in 
module abstract.   
An earlier Java-based client called HotGAMS~\cite{HotGAMS} 
allowed an interactive search of the GAMS repository.  
Both the Survey and GAMS index into Netlib for 
software downloads.   
Another related taxonomy example not from linear algebra is the decision tree for 
optimization 
software at Arizona State University~\cite{optimization-decision-tree}. It 
refers to Netlib entries if available or points directly to the home page of the 
software package. 


\begin{figure*}
\center
\includegraphics[width=\textwidth]{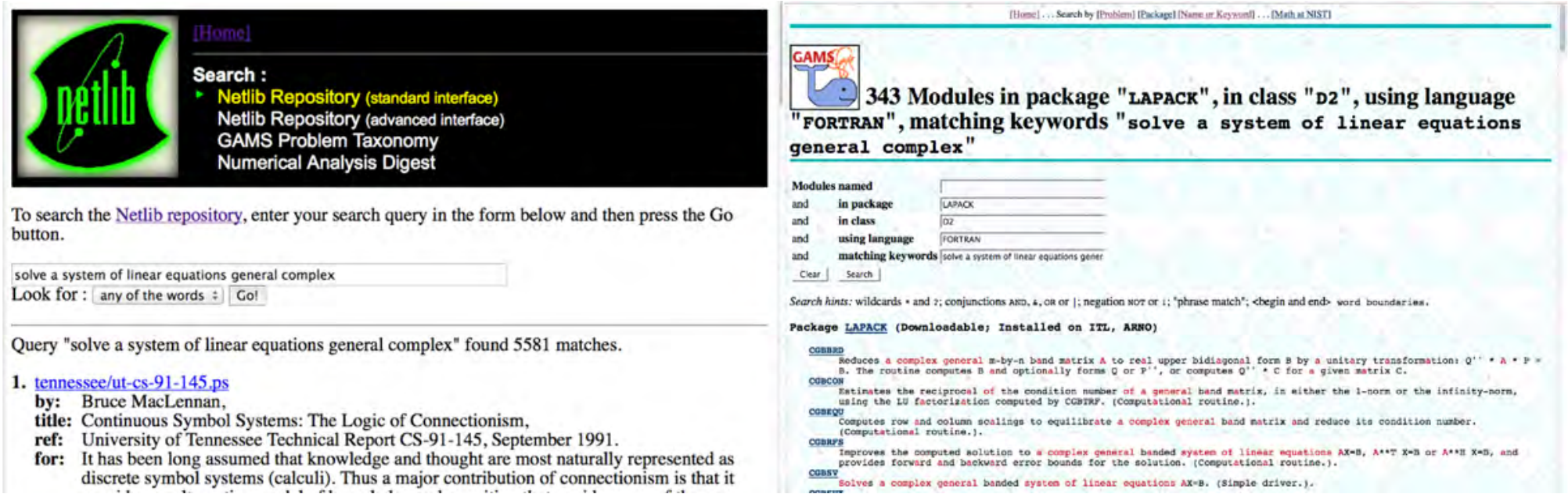}
\vspace{-15pt}
\caption{Keyword search ``solve a system of linear equations general complex'' with Netlib (left) and with GAMS (right).}
\label{fig:taxonomy}
\end{figure*}

While existing taxonomies represent extensive and valuable resources, the function-level indexing approaches they implement are
usually unable to accommodate high-level operations for which no library implementation is available or for which complex software packages,
such as PETSc and Trilinos~\cite{trilinos:Online}, are required.
In addition, they serve primarily as references to numerical software collections and require significant linear algebra knowledge to
use effectively. Searching for desirable solvers with these taxonomies can be difficult, and little information about the routines is typically
returned. For example, as demonstrated in Figure~\ref{fig:taxonomy}, a keyword search 
of the the Netlib Repository (standard interface)~\cite{netlib:Online} for
``solve a system of linear equations general complex'' using
``any of the words'' returns 
5581 results containing routines for various purposes, technical reports, conference
papers, documentation files, and broken link.  Changing this to
``all the words'' results in no matches even though Netlib does include routines for solving systems of linear
equations with dense, complex coefficient matrices.
The same keyword search repeated on NIST's Guide to Available Mathematical Software
(GAMS)~\cite{GAMS} returns 1,485 largely irrelevant routines.
The link for each routine leads to a webpage offering only a small amount of information about that
routine. Google and Bing return 3,700,000 and 6,330,000 results, respectively, mostly to non-software related pages.

A final example is the LAPACK Search Engine~\cite{LAPACKSearch}, which provides a simple way to search the list of LAPACK
routines. A user enters a problem type (e.g., linear equations or orthogonal factorization), problem parameters (e.g., 
real, single precision, banded matrix), and the desired operation (e.g., estimate condition number or factor without pivoting).
The search engine then either displays the code for the most appropriate routine or allows the user to download it as a zip or tar file.
The LAPACK Search Engine provides help with various LAPACK-related language but ultimately delivers only the code itself with no added
information about its use.  

Lighthouse was inspired by the LAPACK Search Engine. Like the LAPACK Search Engine and unlike the alternatives, Lighthouse typically leads
users to a single subroutine. Lighthouse surpasses existing taxonomies in the extent of information available about each
routine in the taxonomy, including automatically extracted documentation and additional documentation added manually.
Unlike most taxonomy efforts, Lighthouse goes beyond functional classification of software and provides a variety of code generation 
and optimization capabilities, making it a multi-purpose tool that both educates developers and automates portions of the 
software design and implementation process for HPC applications.

\section{Lighthouse Design}
\label{sec:method}

Lighthouse is an expert system designed to provide four main types of capabilities:
\textit{(1)} functional ontologies of existing numerical software collections;
\textit{(2)} guided, advanced, and keyword-based search for computational routines in the ontologies; 
\textit{(3)} generation of code templates based on taxonomy search results; and
\textit{(4)} a high-level scripting interface that accepts input in a MATLAB-like language and generates optimized C implementations. 
In the remainder of this section, we describe each of these elements.

\subsection{Software Classification}

A growing number of libraries support linear algebra computations. They include a plethora of specialized 
algorithms and data structures. To identify the best strategy for reaching the most appropriate and best 
performing solution for a specific problem, the user answers a series of
succinct questions via guided search interfaces for each software package. 
In each case, the process of answering these questions corresponds to the traversal of a decision
tree from root to terminal nodes or leaves.  The decision nodes translate into the questions
and the terminal nodes indicate or predict target routines. 
In the remainder of this section we overview the general design and implementation aspects of Lighthouse
for LAPACK.
Both the interface and internal 
representation are designed
to be easily updatable and extensible for new numerical libraries. For example, 
the guided search questions corresponding to internal 
nodes of the decision tree are stored in the database and the web forms in the Lighthouse guided search 
interface are 
dynamically generated from that data. Each subsequent question is selected based on the preceding set of 
responses.



\subsection{Search Capabilities and User Interface}

\begin{figure*}
\centering
\includegraphics[width=\textwidth]{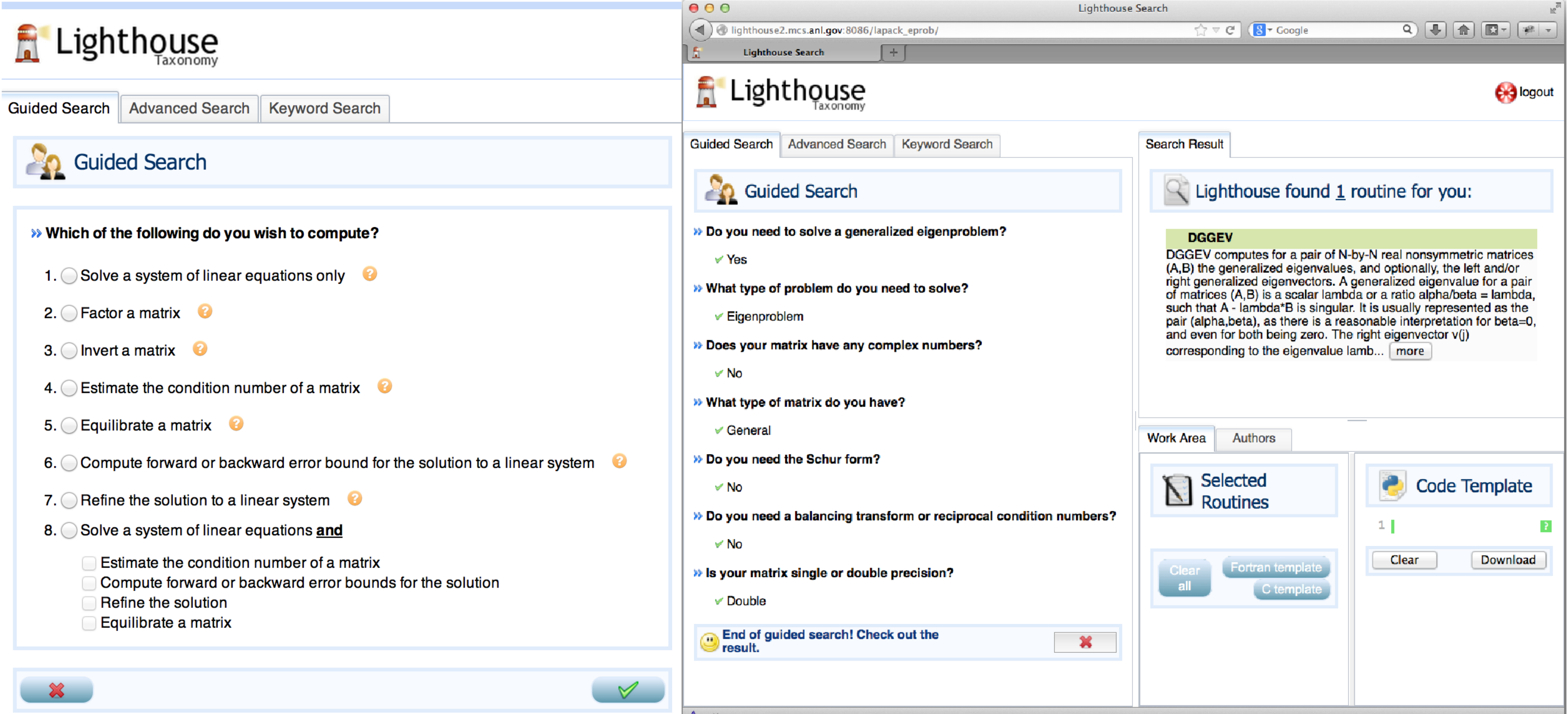}
\vspace{-5pt}
\caption{Guided search dialogs in the Lighthouse LAPACK prototype: The first question of a \emph{linear solver} search (left), and 
the result from a completed \emph{eigen} search that shows all questions and user's answers (right).}
\label{fig:lapackGuided}
\end{figure*}

The Lighthouse taxonomy information is stored in a MySQL database. The main infrastructure of Lighthouse uses Django~\cite{django:Online},
an open-source, high-level Python web framework that provides a dynamic database access application programming interface (API).
To accommodate users with diverse backgrounds and different programming experiences, Lighthouse provides different levels of user interfaces for
subroutine search. For example, through pairing with Haystack~\cite{haystack:Online}, Lighthouse enables numerical 
subroutine search via three methods:
guided search, advanced search, and keyword search. In the \emph{guided search} interface, users are prompted to answer increasingly detailed
questions describing the problems they wish to solve. The response to each question triggers a 
query to the database and directs
the search to the most appropriate result. Portions of the interface are automatically generated to display the responses and search results
using Django dynamic forms and the Django session framework. Figure~\ref{fig:lapackGuided} shows partial guided search dialogs and the
search result. After answering the last question, users typically obtain exactly one routine that matches all of the answers.

Unlike the guided search, the \emph{advanced search} is designed for users who are familiar with the library. The LAPACK advanced search interface
provides users with a form containing checkboxes where they can make multiple selections. The advanced search enables the simultaneous
search for multiple routines in different types of functional categories.

\begin{figure*}[t]
\centering
\includegraphics[width=\textwidth]{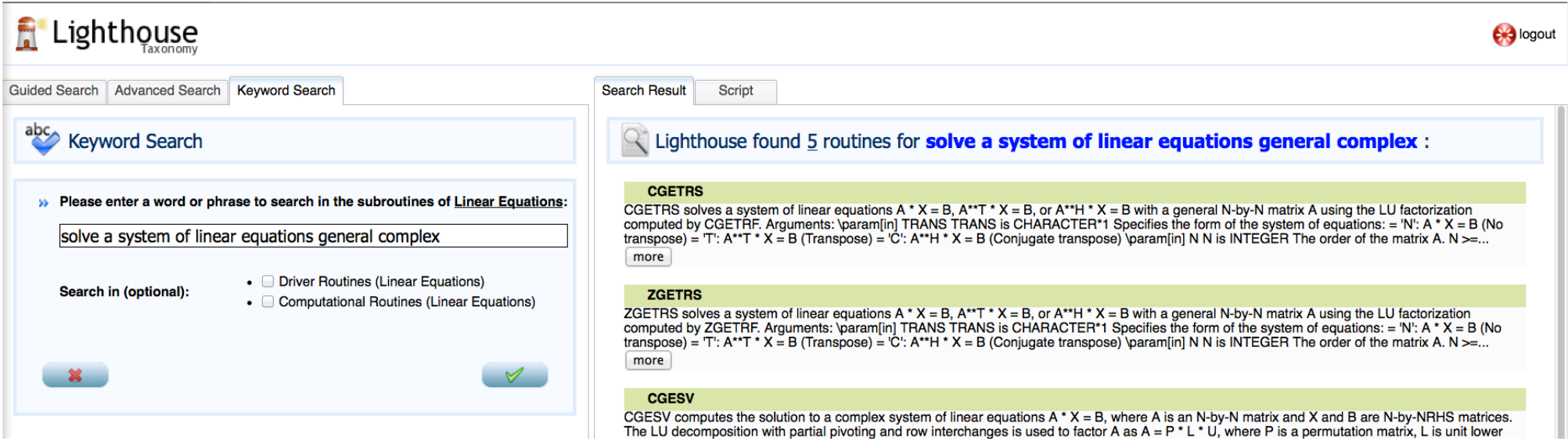}
\vspace{-5pt}
\caption{Keyword search ``solve a system of linear equations general complex'' with Lighthouse for LAPACK.}
\label{fig:lapackKeyword}
\end{figure*}

The \emph{keyword search} interface supports subroutine search based on an input keyword or phrase. 
In order to enhance effectiveness and efficiency, Lighthouse provides automatic completion and spelling correction
with words collected from linear algebra textbook indices. In addition, Lighthouse uses a list of linear algebra terms and phrases for
Django-Haystack filtering in order to reduce search time.
As illustrated in Figure~\ref{fig:lapackKeyword}, Lighthouse returns five LAPACK routines with keywords ``solve a system of linear equations
general complex''.

\subsection{Code Template Generation}

Once a user has identified one or more subroutines through any of the Lighthouse searches, a code template can be generated and downloaded.
The template is a complete compilable and runnable program that declares variables using library-specific data structures, initializes
them properly, and calls the appropriate subroutine(s) correctly.
In addition, each template program is structured into several subprograms, making it easy for users to modify and extend the initial code to fit specific needs.

\begin{figure}[ht]
\centering
\includegraphics[width=.48\textwidth]{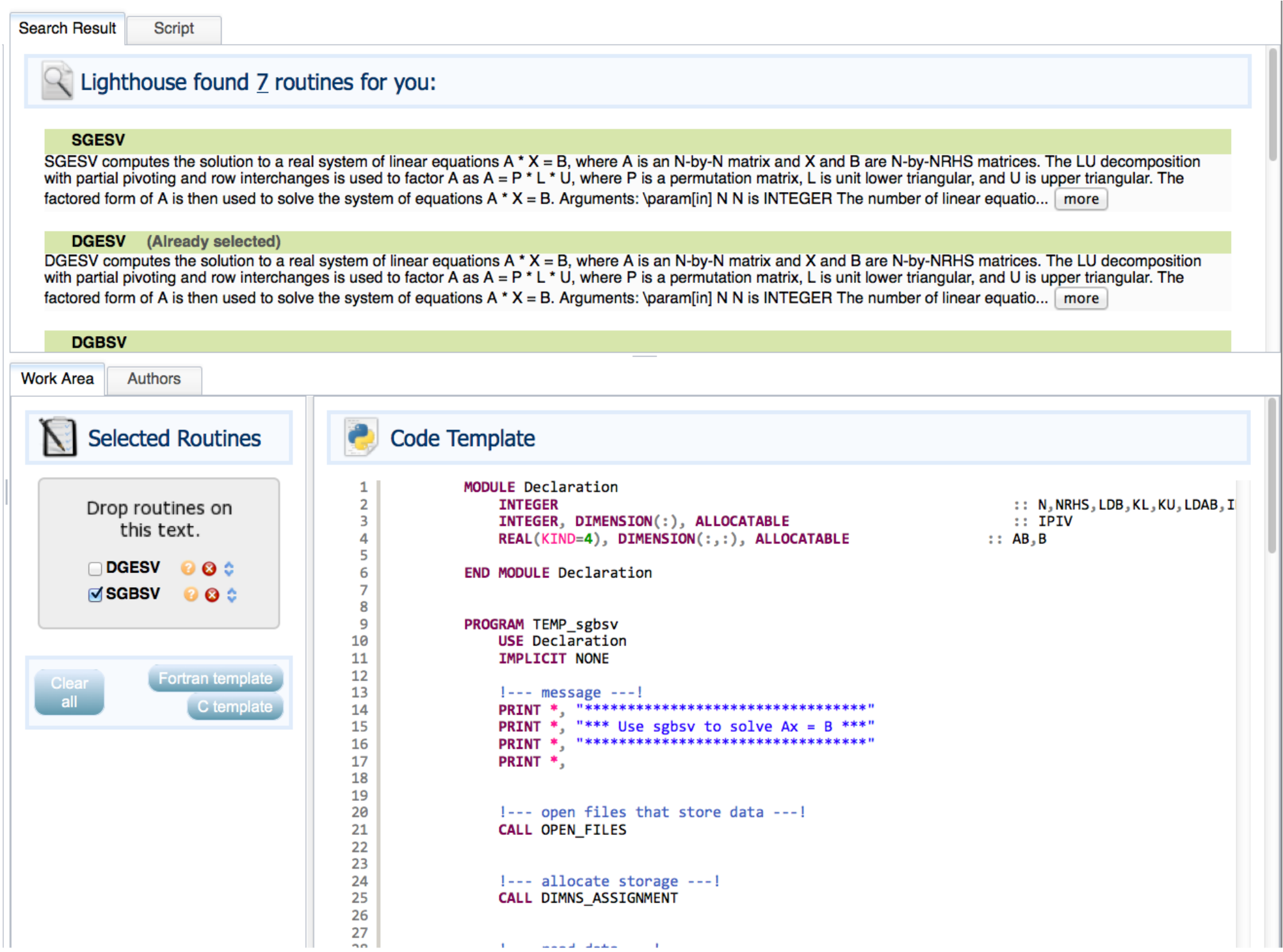}
\caption{Fortran template generated for the LAPACK subroutine DGBSV.}
\label{fig:lapackCode}
\end{figure}

To generate the templates for LAPACK routines, we first constructed a database that stores the parameters of the subroutines and their types.
We then created template \textit{base} files in both Fortran 90 and C based on several different categories of subroutines. 
When a user clicks the template
generation button, appropriate template base files are extended automatically to include the specific parameter information
from the database for the selected subroutine. 
Figure~\ref{fig:lapackCode} illustrates the Fortran code template generated for the LAPACK dense linear system solver DGBSV.

\subsection{Autotuning}

Lighthouse provides preliminary support for generating highly optimized code for custom dense linear algebra computations 
by interfacing with the Build to Order
(BTO)~\cite{Belter,Siek} compiler. BTO's input is a MATLAB-like language for defining sequences of basic linear algebra operations.
The ability to generate custom 
optimized implementation of certain linear algebra computations is useful when there is no existing optimized library implementation.
BTO performs a number of domain-specific optimizations and employs an empirical autotuning approach to generate a high-performance
C implementation of the operations specified by the input.
Lighthouse implements a web-based client and corresponding BTO servers running on remote machines. 
The user inputs the script into the Lighthouse
web interface, which provides syntax checking and type inference (types can also be optionally specified explicitly by the user). After
verifying (at the client) that the syntax is correct, the code is sent to the BTO server, which then generates different low-level
implementations, which 
are compiled and executed on the remote server to determine the best-performing variant.
Because autotuning is most effective when performed 
on a system similar to that which will be used for production application runs, we will eventually support the use of user-selected servers. 
The current prototype supports a limited fixed set of servers managed by us.

\begin{figure}[ht]
\centering
\includegraphics[width=.48\textwidth]{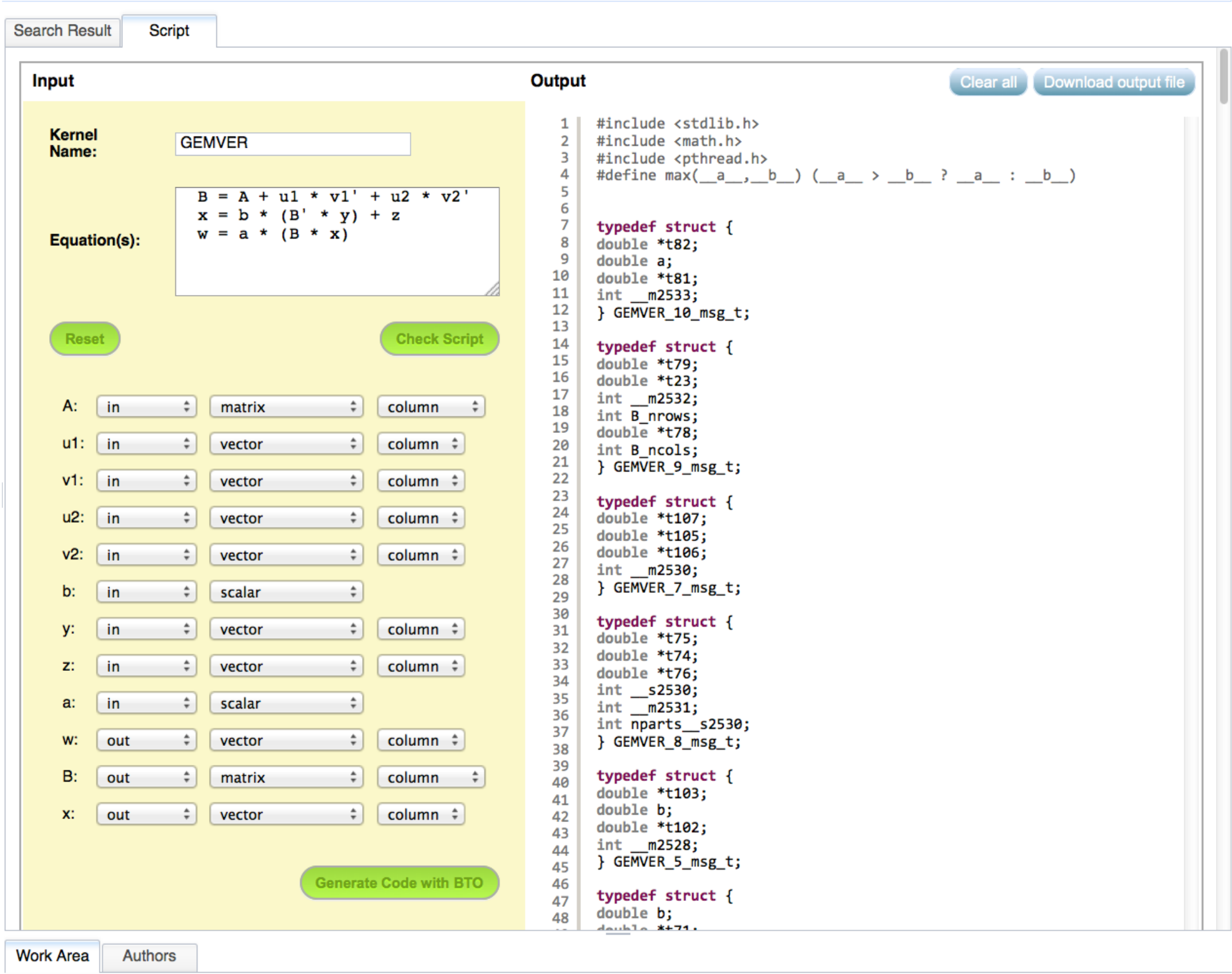}
\vspace{-5pt}
\caption{Script interface with optimized C code generated by BTO.}
\label{fig:bto}
\end{figure}

Figure~\ref{fig:bto} illustrates the use of Lighthouse and BTO to generate and autotune C code for computing the following matrix-vector operations:
\begin{align*}
  B &= A + u_1 * v_1' + u_2 * v_2' \\
  x &= b * (B' * y) + z \\
  w &= a * (B * x),
\end{align*}
where $A$ and $B$ are matrices, $u_1$, $u_2$, $v_1$, $v_2$, $w$, $x$, $y$, and $z$ are vectors, and $a$ and $b$ are scalars.
The ``Input'' area in the Lighthouse script section contains two text fields that allow users to input a kernel name and a sequence
of linear algebra equations in a high-level MATLAB-like syntax. Once the information is submitted, a list of parameters is dynamically
generated including automatic setup of procedure arguments (in, out, inout), data type specifications (scalar, vector, matrix), and
orientation declaration (row, column).  Users can change these pre-declared values by selecting different options from drop-down lists. When the
``Generate Code with BTO'' button is clicked, an input file is automatically created and submitted to the server. After the 
autotuning is complete, the
optimized C code is displayed in the ``Output'' area and can be downloaded by clicking the ``Download output file'' button. Users
can modify and integrate the C code into a larger application context to maximize performance.

\section{HPC Libraries}
\label{sec:result}

We are continually extending Lighthouse by adding new functionality and improving performance. The currently
supported libraries represent an important milestone because they cover a broad space of sequential
and parallel solution methods
for sparse and dense eigenproblems and systems of linear equations.
The Lighthouse application currently offers access to functionality from three linear algebra packages: LAPACK, PETSc, and SLEPc. 
In the remainder of this section we discuss specific use cases and implementation details for each of the three libraries. 

\subsection{Lighthouse for LAPACK}
LAPACK (Linear Algebra Package) is a large Fortran library for numerical linear algebra.
Being one of the most widely used dense direct solver packages for dense matrices, LAPACK was the
logical first choice for inclusion in the Lighthouse taxonomy.
The Lighthouse taxonomy is continuously expanding and presently contains over 800
LAPACK routines for performing a variety of linear algebra tasks.

The data structures used by LAPACK for different kinds of linear systems play an important role in 
the decision trees for selecting appropriate subroutines.
For example, the linear solver subroutines are first categorized by the kinds of tasks they perform. The system then sorts and identifies
eight different matrix types and five different storage methods. Next it categorizes the precision level (single or double) and
parameter type (real or complex) of the routines. The database and the step-by-step questions in the guided search abstracted from the
decision tree form an effective system for helping users make informed decisions about the most desirable routines. The following
use case describes a scenario where a Lighthouse user interacts with the guided search to obtain a single LAPACK linear solver routine.

\begin{enumerate}
\item \textbf{Use Case Name:} Searching for an LAPACK routine for solving a system of linear equations.
\item \textbf{Actors:} User, Lighthouse
\item \textbf{Preconditions:} There is an active network connection to Lighthouse; User is at the guided search page of Lighthouse for LAPACK.
\item \textbf{Basic Flow of Events:}
    \begin{enumerate}
        \item The use case begins when the guided search page for LAPACK is opened.
        \item Lighthouse asks ``Which of the following functions do you wish to execute?'' and displays all the different alternatives that
        are available on this page.
        \item User selects ``Solve a system of linear equations only'' and submits that answer.
        \item Lighthouse displays resultant routines in the Search Result area and asks ``What form of the linear system do you want to solve?''
        \item User selects ``AX = B'' and submits the answer.
        \item Lighthouse updates the search results and asks ``Are there complex numbers in your matrix?''
        \item User selects ``No'' and submits the answer.
        \item Lighthouse updates the search results and asks ``What is the type of your matrix?''
        \item User selects ``general'' and submits the answer.
        \item Lighthouse updates the search results and asks ``How is your matrix stored?''
        \item User selects ``band'' and submits that answer.
        \item Lighthouse updates the search results and asks ``Would you like to use single or double precision?''
        \item User selects ``double'' and submits the answer.
        \item Lighthouse returns the final result, DGBSV, and displays ``End of guided search! Check out the result.''
        \item The use case ends successfully.
    \end{enumerate}
\item \textbf{Post-conditions:} User has obtained the target LAPACK linear solver routine  
\end{enumerate}

\begin{figure}[ht]
\centering
\includegraphics[width=.48\textwidth]{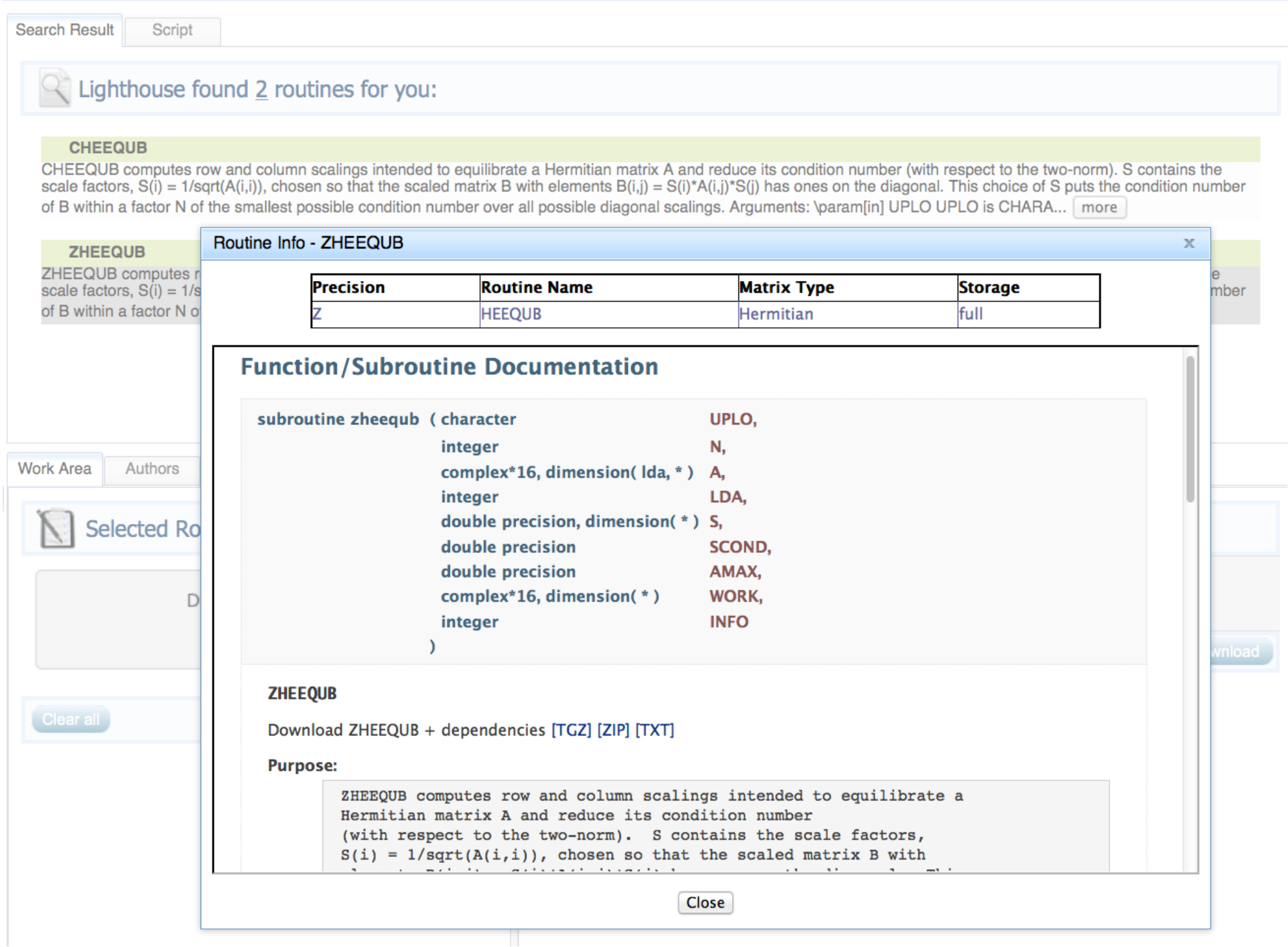}
\vspace{-5pt}
\caption{A pop-up window displaying subroutine documentation.}
\label{fig:document}
\end{figure}

In addition to providing the search and code generation capabilities described in Sec.~\ref{sec:method}, Lighthouse serves as an educational tool.
Information about the definition of a word or a phrase, the numerical library, and individual subroutines can be easily accessed by clicking on the help
icons or the ``more'' buttons available throughout the page. Figure~\ref{fig:document} shows a pop-up window containing
detailed information for a particular subroutine. The intuitive drag-and-drop feature within the Lighthouse interface
enables users to simply select a routine by grabbing it and dragging it to the ``Selected Routines'' work area for code template
generation as shown in Figure~\ref{fig:lapackCode}.

Several of the Lighthouse LAPACK code templates contain decision logic to enable tailoring of the solution to specific problem characteristics.
For example, code templates for equilibration routines can also determine whether a given matrix is worth scaling. If it is, a method
of matrix scaling (row, column, or row-and-column) is automatically selected and executed, and the scaled matrix is returned.
The downloadable LAPACK code templates are stored archived for downloading. Each zip file contains a code template file, a README file, and a portable
makefile for building on most unix platforms.
The README file explains in detail how to best use the template. The makefile helps users compile the code and can be easily modified to
suit users' specific needs.

\subsection{Lighthouse for PETSc}

For solving large systems of sparse linear equations, Lighthouse provides search and code generation capabilities for the widely used
Portable Extensible Toolkit for Scientific Computation (PETSc)~\cite{petsc:Online}. 
Lighthouse for PETSc has a simple and yet efficient navigation system that enables users to generate, download
and extend complete PETSc-based applications for solving large sparse systems. 
Specifically, it automates the nontrivial selection of a Krylov subspace (KSP) method and a preconditioner to accelerate
convergence rate and reduce execution time. What makes linear solver method selection difficult in practice is that 
different methods can exhibit very different convergence behavior depending on the (possibly unknown) properties of the linear system
and, in many cases, may fail to converge at all. At the time of this writing, PETSc provides more than 20 Krylov subspace methods and over 
35 preconditioning methods. The task of selecting ``the best'' solver-preconditioner combination among the hundreds or thousands of valid pairings
and configuring them in a way that maximizes performance is currently left to the application developer who is typically 
a scientist or engineer whose primary expertise is not in applied mathematics or HPC algorithms. Moreover, for most of these methods, 
one cannot analytically determine the best algorithms even if all linear system properties are known in advance.

The integration of Lighthouse aims to address two different needs. First, just as they do for dense linear system problems, users interactively
express their problem in the way most natural to them, and, based on the user input, Lighthouse generates a complete code template
and a set of configuration options that can be used as a starting point for future extensions. Second, Lighthouse tackles the problem
of determining what ``the best'' solution method (KSP solver and preconditioner pair) is for a given linear system. Users have the option
of either uploading a sample linear system for Lighthouse to analyze, or they can specify the properties they know directly.

To illustrate how Lighthouse supports solving large sparse linear systems by using PETSc, we describe an example use case.
\begin{enumerate}
\item \textbf{Use Case Name:} Solving a large sparse linear system of equations. 
\item \textbf{Actors:} User, Lighthouse
\item \textbf{Preconditions:} There is an active network connection to Lighthouse; the user is at the guided search page of Lighthouse for PETSc.
\item \textbf{Normal flow:}
    \begin{enumerate}
        \item Lighthouse asks the user if they want to upload their matrix.
	\item User chooses to upload their matrix.
	\item Lighthouse provides the user with a file browser for selecting the matrix file.
	\item User selects the matrix file.
	\item Lighthouse asks the user if they want a sequential solution or a parallel solution.
	\item User selects the type of solution they want.
	\item User submits the form.
	\item Lighthouse generates a PETSc program in C programming language, a makefile, a text file containing the command line options and a README file.
	\item Lighthouse then creates an archive file containing the files generated in the previous step.
	\item Lighthouse provides the user with a download link of the archive file.
	\item User downloads the archive file.
	\item Lighthouse deletes user files from the server.
        \item The use case ends successfully.
    \end{enumerate}
\item \textbf{Post-conditions:} User has obtained an application template and a set of options for configuring the best sparse linear solver and preconditioner for the given linear system.
\end{enumerate}
\subsubsection{PETSc Search Interface}
To implement the
PETSc guided search, we began by categorizing matrices based on matrix properties including structural (e.g., symmetry, average or max/min number
of nonzeros per row), norm (e.g., 1-, infinity-, and Frobenius norms), spectral (e.g., condition number estimates), normality, and
variance (e.g., element variability in rows and columns). 
Next, we considered a number of use cases~\cite{uml:Online} to organize and model the interaction
between users and Lighthouse. The resultant data set and use cases have suggested that Lighthouse for PETSc provides users with
the following options of operation:

\begin{enumerate}
    \item Compute matrix properties with Lighthouse;
    \item Download a PETSc program for computing matrix properties;
    \item Download a general PETSc program for solving a linear system;
    \item Upload matrix properties computed using the user's own program.
\end{enumerate}

\begin{figure}[ht]
\centering
\includegraphics[width=.47\textwidth]{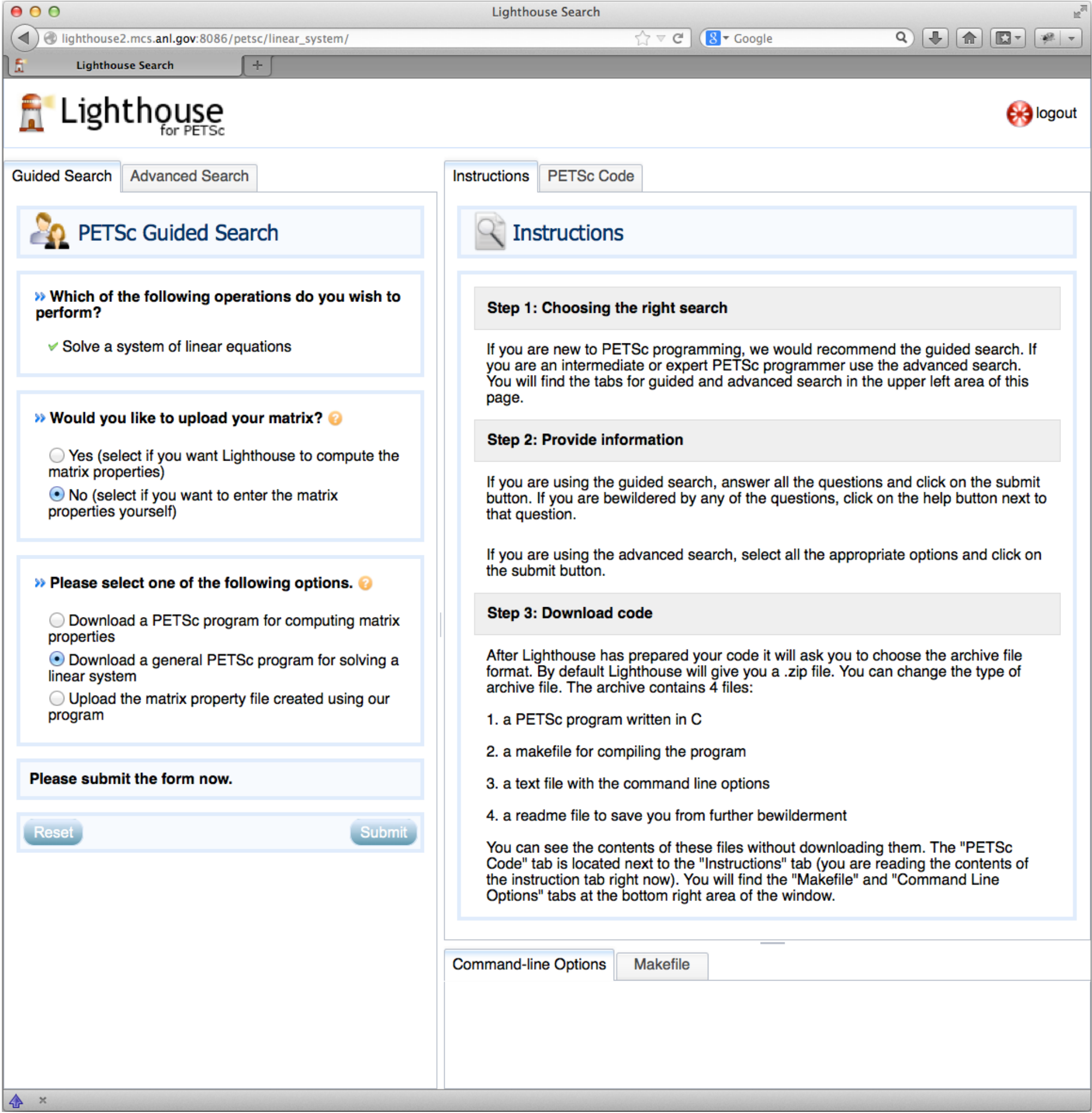}
\vspace{-5pt}
\caption{Lighthouse guided search prototype for PETSc.}
\label{fig:petsc}
\end{figure}

\begin{figure}[htb]
\centering
\includegraphics[width=.47\textwidth]{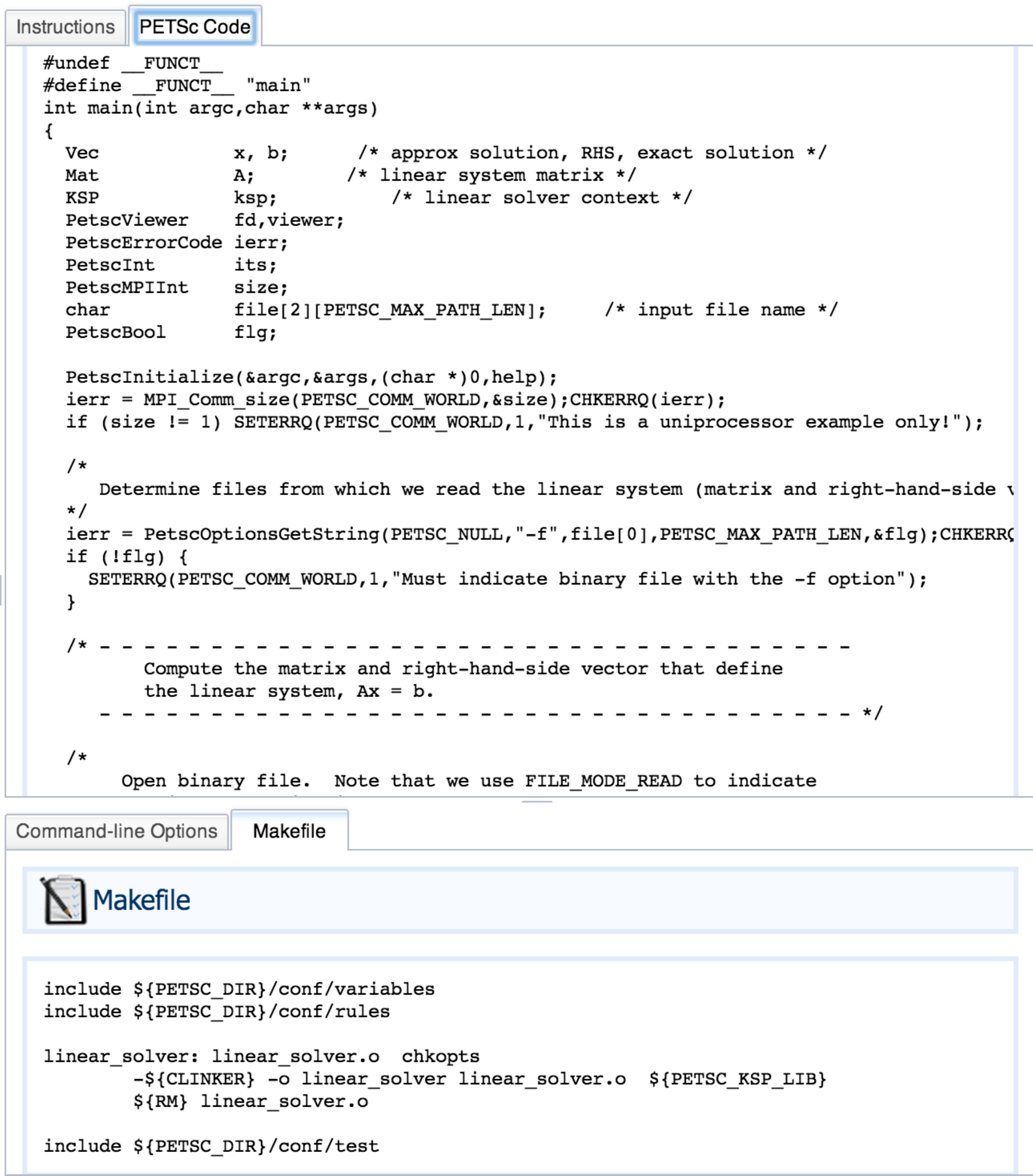}
\vspace{-5pt}
\caption{PETSc search result: Code template and Makefile.}
\label{fig:makefile}
\end{figure}

If option (1) is selected, Lighthouse provides code for the task and a makefile for compiling the code, both of which can be viewed in the
respective tabs before downloading. If option (2) is selected, an archive file containing the matrix property
computation program and other necessary files is provided. If option (3) is selected as shown in Figure~\ref{fig:petsc}, a default
PETSc program without a suggestion of a set of a Krylov subspace method and a preconditioner is generated
along with other necessary files. Figure~\ref{fig:makefile} illustrates portions of the resulting code template and makefile.
This option is for users who are familiar with Krylov methods and know the preconditioner they wish to use. If option (4) is selected,
users are guided to answer increasingly detailed questions, and an archive file that includes the program and the files required for the
program is delivered at the end of the process.

\subsubsection{Solver Selection}
To select a well-performing method, we must have a way to predict its performance for any valid input. In the case of sparse linear solvers,
the inputs are the linear system (matrix) and one or more right-hand side vectors. All of the Krylov subspace methods in PETSc have the
same asymptotic complexity but can differ significantly in their convergence rates for a given linear system. 
Moreover, the choice of preconditioning method is 
also very important (there is a tradeoff between computational cost and the accuracy of the inverse approximation produced by a given preconditioner).
Because of these complexities (and lack of theoretical models), we modeled the performance of solver/preconditioner methods by using a 
machine learning approach.

First, we considered a set of 30 linear system properties, including 
matrix dimension, nonzero statistics, symmetry and other structural characteristics, trace, various norms, dominance, and variance. Because such 
a large set of features is too expensive to compute during the application's execution, we applied principal component analysis 
(PCA)~\cite{pca:Online} to the  data generated by using different solvers to solve 860 real square linear systems
obtained from the University of Florida Sparse Matrix Collection~\cite{UofF:Online}. We solved each of these sparse linear system by using
different combinations of a Krylov subspace method (including CG, CGS, BICG, BICGSTAB, GMRES, FGMRES, and TFQMR)
and a preconditioner (including ILU(k) with different levels of fill, Jacobi, block Jacobi, SOR, and ASM), 
and evaluated the convergence and the time to solution for each
computation. 
Then we applied PCA by using the MATLAB Statistics Toolbox
~\cite{MATLABstats} 
and identified 15 properties that account for the greatest variability in the data, which enabled us to reduce the feature set from 30 properties to 15 
(shown in Table~\ref{tab:featureset}). For very long-running simulations, using better-performing solvers quickly amortizes the initial 
overhead for computing the features needed for selecting the solver and preconditioner.

\begin{table}[htb]
 \centering
    \caption{Feature set for PETSc linear solver classification and average computation time for calculating each feature for matrices in our training set.}
   \label{tab:featureset}
    \begin{tabular}{| l | c |}
    \hline
    \multicolumn{2}{|c|}{Reduced feature set and computation time} \\
    \hline
    \hline
    Matrix feature & Computation time \\ \hline
    Row variance & 9.90s \\ \hline
    Column variance & 11.47s \\ \hline
    Diagonal variance & 0.16s  \\ \hline
    Number of nonzeros & 1.97s  \\ \hline
    Number of rows & 0.06s \\ \hline
    Frobenius norm & 0.03s  \\ \hline
    Symmetric Frobenius norm & 13.34s  \\ \hline
    Anti-symmetric Frobenius norm & 13.24s  \\ \hline
    One norm &  0.21s \\ \hline
    Infinity norm & 0.07s \\ \hline
    Symmetric infinity norm & 13.33s \\ \hline
    Anti-symmetric infinity norm & 13.44s \\ \hline
    Max. nonzeros per row & 1.95s \\ \hline
    Trace & 0.06s  \\ \hline
    Absolute Trace & 0.15s \\ \hline
    \hline
    Total time & 79.38s \\ \hline
   \end{tabular}
\end{table}

Second, we explored different machine learning techniques for building a model of the performance based on the set of linear system 
features, including Naive Bayes classifiers, support vector machines (SVM), alternating decision trees, decision stumps, and instance-based
learning. For our training set, SVM~\cite{svm} was consistently more accurate than the other approaches in 
selecting the best combination of a Krylov subspace method and a preconditioner for any inputs (not in the training set). 
The average accuracy of the prediction for sequential problems is 85.41\% and for parallel problems is 81.95\%. As users contribute 
more linear systems, we will periodically update our training set and update the models to improve the accuracy of 
the classification for a wider variety of linear systems.

At present we use the classifier to create a single static configuration for a given linear system. In many applications, such as 
the solution on nonlinear partial differential equations through Newton-Krylov methods, one must solve many sparse linear systems during
a single application execution. While some of the linear system properties may remain the same throughout such simulations, others 
may change, and the initial choice of Krylov method/preconditioner pair may not be the best for all linear systems 
that arise in a single application run. We plan to integrate some of the adaptive linear solver techniques we have explored 
previously~\cite{Raghavan:2005:IPDPS,Bhowmick:2004:ASTC} 
into Lighthouse by generating code templates that allow runtime method 
selection by periodically computing the properties of selected linear systems and applying the ML-based model to select 
a different solver/preconditioner combination.

\subsection{Lighthouse for SLEPc}
Built on top of PETSc, SLEPc is a parallel toolkit for solving large sparse eigenproblems and is the most recent addition to the
Lighthouse taxonomy. SLEPc provides six main types of solvers: EPS (Eigenvalue Problem Solver), ST (Spectral Transform),
SVD (Singular Value Decomposition), QEP (Quadratic Eigenvalue Problem), NEP (Nonlinear Eigenvalue Problem), and MFN (Matrix Function).
Our work has primarily focused on the analysis of the EPS eigensolver routines and their integration into Lighthouse. 

To illustrate how Lighthouse supports solving large sparse eigensystems by using SLEPc, we describe an example use case.
\begin{enumerate}
\item \textbf{Use Case Name:} Searching the SLEPc package for the best eigensolver routine for a sparse matrix.
\item \textbf{Actors:} User, Lighthouse
\item \textbf{Preconditions:} There is an active network connection to Lighthouse; the user is at the guided search page of Lighthouse for SLEPc.
\item \textbf{Basic Flow of Events:}
    \begin{enumerate}
        \item The guided search page for SLEPc is opened and the series of questions corresponding to the features of the matrix and the result characteristics are displayed.
        \item User answers questions with respect to matrix properties like Hermitian or non-Hermitian, real or complex, etc.
        \item User selects expected properties of the results such as the number of eigenvalues, tolerance desired, portion of eigenvalue spectrum.
	\item User selects available resources such as number of processors available.
        \item Every user selection updates the results matching solvers (e.g., Power, Arnoldi).
        \item Lighthouse returns the final result of best solvers when all questions are answered.
        \item The use case ends successfully.
     \end{enumerate}
\item \textbf{Post-conditions:} User has obtained the best sparse eigensolver routine(s) with respect to the conditions specified.
\end{enumerate}

Most of the eigensolvers available in SLEPc are compatible with most types of problems. However, not all
solvers are equal in terms of accuracy and efficiency. An improper solver selection can result in an unacceptably 
high residual, lead to
essentially no convergence, or yield no feasible solution. By analyzing SLEPc solvers for various input conditions and applying
machine learning techniques to the resultant data, we have enabled Lighthouse to identify reliably performant methods
for finding eigenvalues and eigenvectors of wide variety of input problems.

Unlike LAPACK linear solvers, where typically there is one solution method that is best suited to a particular problem, 
there can be multiple SLEPc eigensolver routines providing similar performance and accuracy. Hence, instead 
of finding a single best solver we look for multiple SLEPc routines that exhibit similar performance for different input cases.
To determine the best SLEPc eigensolvers for a given problem, we experimented with compatible routines for various
matrices and input parameters using matrices obtained from 
the University of Florida Sparse Matrix Collection and Matrix Market~\cite{matrixMarket:Online}. 
Table~\ref{table_exp} shows the range of input cases
tested, and Table~\ref{table_solvers} shows the SLEPc solvers run for these input cases.
We recorded the statistics of the output characteristics including time to solution, residual, and number of converged 
eigenvalues. Some of the output characteristics vary substantially with different solvers
for the same matrix and input parameters.
\\
\begin{table}[!htbp]
\centering
\caption{Input parameters used for eigensolver classification.}
\label{table_exp}
\begin{tabular}{|l|p{4.5cm}|}
\hline
\textbf{Property tested}                    & \textbf{Range of sample tested}                                                                                                 \\ \hline
Matrix order            & $112\times112$ to $262111\times262111$                                                                            \\ \hline
Matrix type           & Real, Complex
\\ \hline
Matrix data           & Binary , Double
\\ \hline
Matrix characteristics & Hermitian, Non-Hermitian
\\ \hline
Number of eigenvalues  & 1,2,5,10
\\ \hline
Portion of spectrum    & Largest magnitude, Smallest magnitude, Largest real, Smallest real, Largest imaginary, Smallest imaginary 
\\ \hline
Tolerance              &  1.00E-04, 1.00E-08, 1.00E-10
\\ \hline
Number of processors   & 1, 2, 4, 8, 12, 24, 48, 96, 192                                                                            \\ \hline
\end{tabular}
\end{table}

\begin{table}[!htbp]
\centering
\caption{SLEPc solvers in the taxonomy.}
\label{table_solvers}
\begin{tabular}{|c|p{5.3cm}|}
\hline
\textbf{SLEPc solvers} & power, subspace, arnoldi, lanczos, krylovschur, generalized davidson, jacobi davidson\\ \hline
\end{tabular}
\end{table}

We collected over 29,000 data points and analyzed them with MATLAB to first remove solvers 
whose number of converged 
eigenvalues was below the
desired value or whose tolerance for the residual was above the threshold. This step gave us a set of eligible solvers. The eigensolver that
took the least amount of time was selected as the best fit. In addition, 
we also identified eigensolvers that take less than $110\%$ of the time taken by the best in order
to provide to the users a collection of eigensolvers with similar performance. In 10-fold cross-validation, the average accuracy 
of predicting the best performing solvers(s) was 86.39\%.

\begin{figure}[ht]
\centering
\includegraphics[width=.48\textwidth]{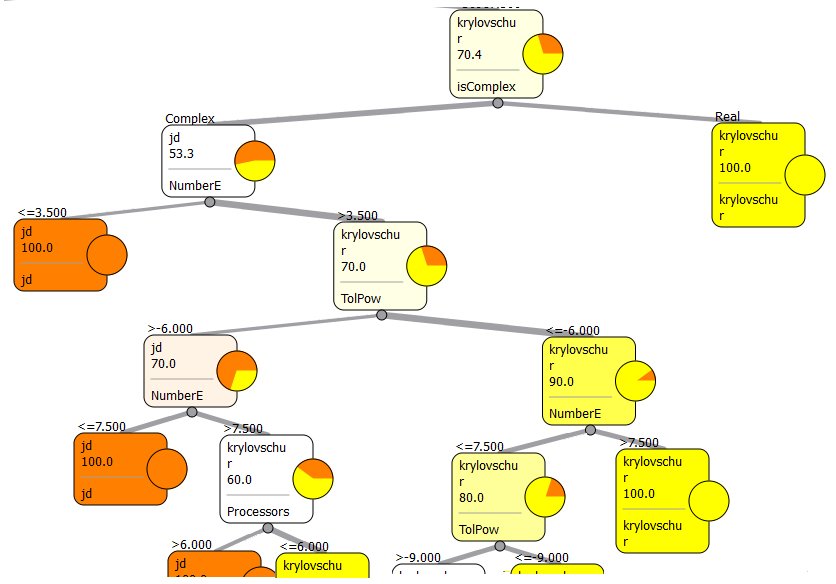}
\vspace{-5pt}
\caption{A portion of the decision tree for SLEPc.}
\label{fig:DecisionTree}
\end{figure}

These data formed a training set for machine learning to create a model that can 
make intelligent decisions when selecting solvers for
any untested cases. We used MATLAB to implement decision tree induction and adopted ORANGE~\cite{orange:Online},
an open source data mining tool, for generating a classifier in the form of a binary tree as shown in Figure~\ref{fig:DecisionTree}.
Each node of the tree corresponds to one of the input features such as matrix size, matrix type, and desired eigenvalue spectrum.

To integrate the classification tree information into the Lighthouse interface, we converted the decision tree into a MySQL 
data table using a depth first search algorithm~\cite{Skiena:1997}. Every path from the root node to the leaf node forms a 
single row in the data table. The 395 leaf nodes in the generated
decision tree thus resulted in 395 distinct rows with unique features. 
The user selects the desired features via the Lighthouse-SLEPc interface and the respective best solvers are fetched from the database.
The ability to automatically update the decision tree based on new training data ensures that the SLEPc 
solver recommendations can be improved  

\begin{figure}[ht]
\centering
\includegraphics[width=.48\textwidth]{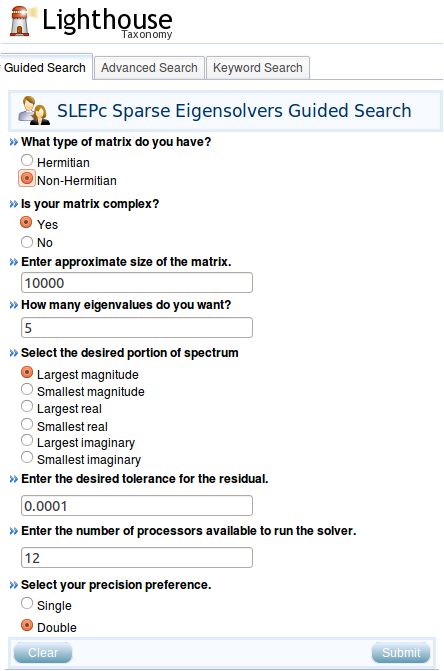}
\vspace{-5pt}
\caption{Guided search interface for sparse eigensolvers.}
\label{fig:SLEPc_GS}
\end{figure}

The classification information obtained by using extensive experimentation and machine learning 
enables Lighthouse to predict the best SLEPc eigensolver for given input and output conditions of
a problem. In addition, it provides insight into the design of the guided- and advanced-search interfaces of Lighthouse for SLEPc.
Figure \ref{fig:SLEPc_GS} illustrates the guided search user interface, consists of a series of simple
questions. Answering these questions retrieves the SLEPc eigensolver that works best for the given conditions.

\section{Conclusion and Future Work}
\label{sec:conclusion}

This paper describes our work on the development of Lighthouse, a novel user-centered web service for numerical linear algebra software.
The Lighthouse framework comprises a prototype taxonomy of available collections of high-per\-formance 
numerical software currently including LAPACK, PETSc, and SLEPc.
Not only does Lighthouse offer more accurate and robust search capabilities than do the existing taxonomies, it also provides code templates and
easy access to the BTO compiler for generating highly tuned implementations of fused linear algebra operations.  By following
the levels of the decision tree, it identifies the best choice of LAPACK routine for dense linear systems. By applying machine learning techniques,
Lighthouse is able to suggest a combination of a Krylov subspace method and a preconditioner for a PETSc linear solver with the best expected
performance or a
set of SLEPc eigensolvers with similar expected performance.

In future work, we will continue to add linear algebra functionality to Lighthouse.  In progress now are the inclusion of
dense eigensolvers and orthogonal factorizations from LAPACK and sparse linear systems from PETSc.
We will also
 continue to build a complete user interface for each of the existing libraries in Lighthouse. We will also 
expand the taxonomy and the autotuning support for new types of numerical computations. 
To take Lighthouse to the next level, we intend to automate the collection
and use of performance information in taxonomy searches. We will ultimately provide innovative methods for building a communication platform
for cross-domain information exchange on topics related to the contents of the taxonomy.

\section*{Acknowledgments}

The authors would like to thank Javed Hossain, Luke Groeninger and Li-Yin Young for their participation and contributions to the project.
This work is funded by the National Science Foundation (NSF) grants (award numbers CCF-0916474 and CCF-1219089) and
utilizes the Janus supercomputer, which is supported by the NSF (award number CNS-0821794) and by the University of Colorado Boulder.
The Janus supercomputer is a joint effort of the University of Colorado Boulder, the University of Colorado Denver, and the National Center
for Atmospheric Research. Janus is operated by the University of Colorado Boulder.

\bibliographystyle{abbrv}
\bibliography{paper}  

\end{document}